\documentclass[p3,sort&compress]{elsarticle}
\usepackage[T1]{fontenc}
\usepackage[latin9]{inputenc}
\usepackage{geometry}
\usepackage{color}
\usepackage{amsmath}
\usepackage{amssymb}
\usepackage{graphicx}
\usepackage{esint}
\usepackage[unicode=true,
 bookmarks=true,bookmarksnumbered=false,bookmarksopen=false,
 breaklinks=false,pdfborder={0 0 1},backref=false,colorlinks=true]
 {hyperref}
\usepackage{breakurl}
\begin{document}

\begin{frontmatter}{}

\title{Steady quantum coherence in non-equilibrium environment}

\author{Sheng-Wen Li}

\author{C. Y. Cai}

\author{C. P. Sun}

\address{Beijing Computational Science Research Center, Beijing 100084, China}

\address{Synergetic Innovation Center of Quantum Information and Quantum Physics,
University of Science and Technology of China, Hefei, Anhui 230026,
China}
\begin{abstract}
We study the steady state of a three-level system in contact with
a non-equilibrium environment, which is composed of two independent
heat baths at different temperatures. We derive a master equation
to describe the non-equilibrium process of the system. For the three
level systems with two dipole transitions, i.e., the $\Lambda$-type
and V-type, we find that the interferences of two transitions in a
non-equilibrium environment can give rise to non-vanishing steady
quantum coherence, namely, there exist non-zero off-diagonal terms
in the steady state density matrix (in the energy representation).
Moreover, the non-vanishing off-diagonal terms increase with the temperature
difference of the two heat baths. Such interferences of the transitions
were usually omitted by secular approximation, for it was usually
believed that they only take effect in short time behavior and do
not affect the steady state. Here we show that, in non-equilibrium
systems, such omission would lead to the neglect of the steady quantum
coherence.\end{abstract}

\end{frontmatter}{}

\section{Introduction}

Isolated quantum systems evolve unitarily according to the Schr\"odinger
equation, while an open quantum system, which is inevitably coupled
to a heat bath in practical, usually quickly lose all its quantum
coherence. That is, all the off-diagonal terms of the density matrix
of the system $\langle E_{m}|\rho|E_{n}\rangle$ (in the energy representation)
will decay to zero when the open system approaches the steady state
\cite{breuer_theory_2002,zurek_decoherence_2003,quan_quantum-classical_2006,li_synchro-thermalization_2014}.
This phenomenon is called decoherence, and it is also believed that
this is why our world appears as a classical one and no macroscopic
superposition can exist stably in usual cases \cite{zurek_decoherence_2008}.
It has been reported that if some non-vanishing steady quantum coherence
exists in certain special environment, even with a quite small amount,
it can result to some novel physics, such as lasing without inversion
\cite{scully_degenerate_1989}, or extracting work from a single heat
bath \cite{scully_extracting_2003,quan_quantum-classical_2006,de_liberato_carnots_2011}.

\begin{figure}
\begin{centering}
\includegraphics[width=0.38\textwidth]{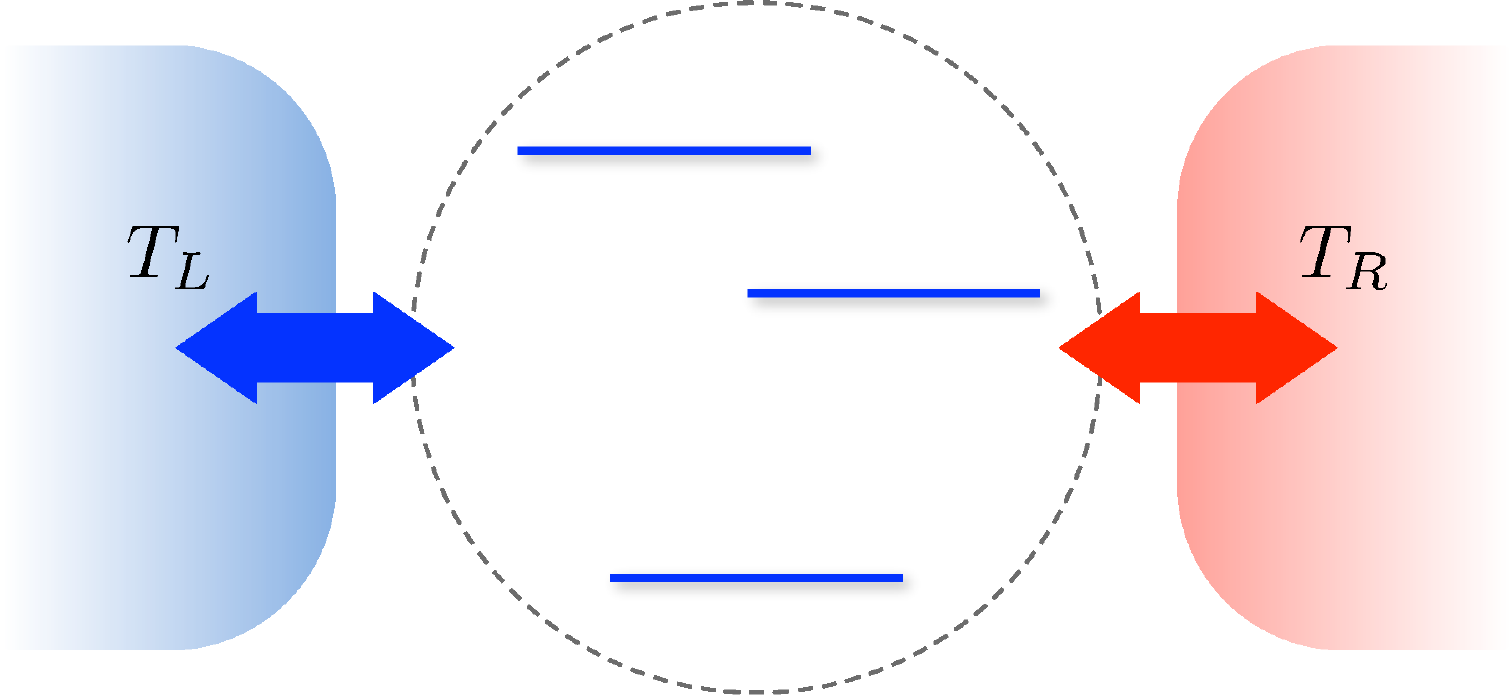}
\par\end{centering}

\protect\caption{(Color online) Demonstration for a non-equilibrium system: a multi-level
system contacted with two heat baths at different temperatures $T_{L/R}$.}

\label{fig-NE}
\end{figure}

Then an important question arises: how can quantum coherence survive
stably in the steady state against decoherence \cite{quan_quantum-classical_2006}?
In this paper, we find that the steady quantum coherence can indeed
exist stably when the system contacts with a non-equilibrium environment,
which is composed of multiple equilibrium heat baths at different
temperatures. Here we study the steady state of a three-level system,
which is coupled to two heat baths with temperatures $T_{L/R}$ respectively
(Fig.\,\ref{fig-NE}). We find that, for the $\Lambda$-type and
V-type systems, non-vanishing quantum coherence can exist in the steady
state when the temperatures of the two heat baths are different. Moreover,
the amount of the nonzero off-diagonal terms increase with the temperature
difference $\Delta T$ of the two heat baths. While the quantum coherence
always vanishes in a $\Xi$-type system. Here we must emphasize that,
unlike previous studies \cite{scully_extracting_2003,de_liberato_carnots_2011},
in our model there is no quantum coherence in the environment in priori,
and the steady quantum coherence in the system is naturally brought
in by the non-equilibrium environment.

Physically, this steady quantum coherence results from the interference
of transitions in non-equilibrium systems. In the three kinds of three-level
systems we study, there are two transition pathways, and there exist
interferences between the transitions for the $\Lambda$-type and
V-type systems \cite{cardimona_steady-state_1982,zhu_quantum-mechanical_1995,zhou_quantum_1997,scully_quantum_1997,kozlov_inducing_2006,kiffner_chapter_2010,tscherbul_long-lived_2014,tscherbul_partial_2015}.
We have to point out that such interferences were often omitted by
secular approximations in many previous literatures \cite{breuer_theory_2002,wu_quantum_2011,vogl_criticality_2012,yin_nonequilibrium_2014}.
Here we show that this omission is consistent for equilibrium environments,
i.e., when all the temperatures of different baths equal to each other
and thus they become a whole equilibrium heat bath. However, in non-equilibrium
systems, such omission of interference between transitions would lead
to the neglect of the steady quantum coherence, and that would also
lead to some other unphysical results \cite{wichterich_modeling_2007,li_long-term_2014}.

Moreover, in a simple example we will show that the quantum coherence
exactly reflects the non-equilibrium flux inside a composite system,
thus it has a clear physical meaning and should not be neglected in
non-equilibrium systems \cite{li_long-term_2014}.

The paper is organized as follows. In Sec.\,II, we derive a master
equation for a $\Lambda$-type system contacting with two heat baths,
and discuss the effect of secular approximation. In Sec.\,III, we
show that non-vanishing quantum coherence can exist in the non-equilibrium
steady state of the $\Lambda$-type system, if we take into account
the interferences between transitions and do not apply the secular
approximation. We also give the condition for the existence of steady
quantum coherence. In Sec.\,IV, we show that steady quantum coherence
can also appear in V-type systems, but cannot appear in $\Xi$-type
systems. In Sec.\,V, we show that the quantum coherence reflects
the non-equilibrium flux inside a composite system. Finally we draw
conclusions in Sec.\,VI.

\section{Non-equilibrium $\Lambda$-type system}

We first consider a $\Lambda$-type system contacting with two independent
heat baths (bath-$L/R$) with different temperatures $T_{L/R}$. We
derive a master equation via Born-Markovian approximation to describe
the dynamics of the open quantum system. Especially, we consider what
physical process has been ignored in the conventional secular approximation
during the derivation of the master equation.

\subsection{Model setup and master equation}

\begin{figure}
\begin{centering}
\includegraphics[width=0.6\textwidth]{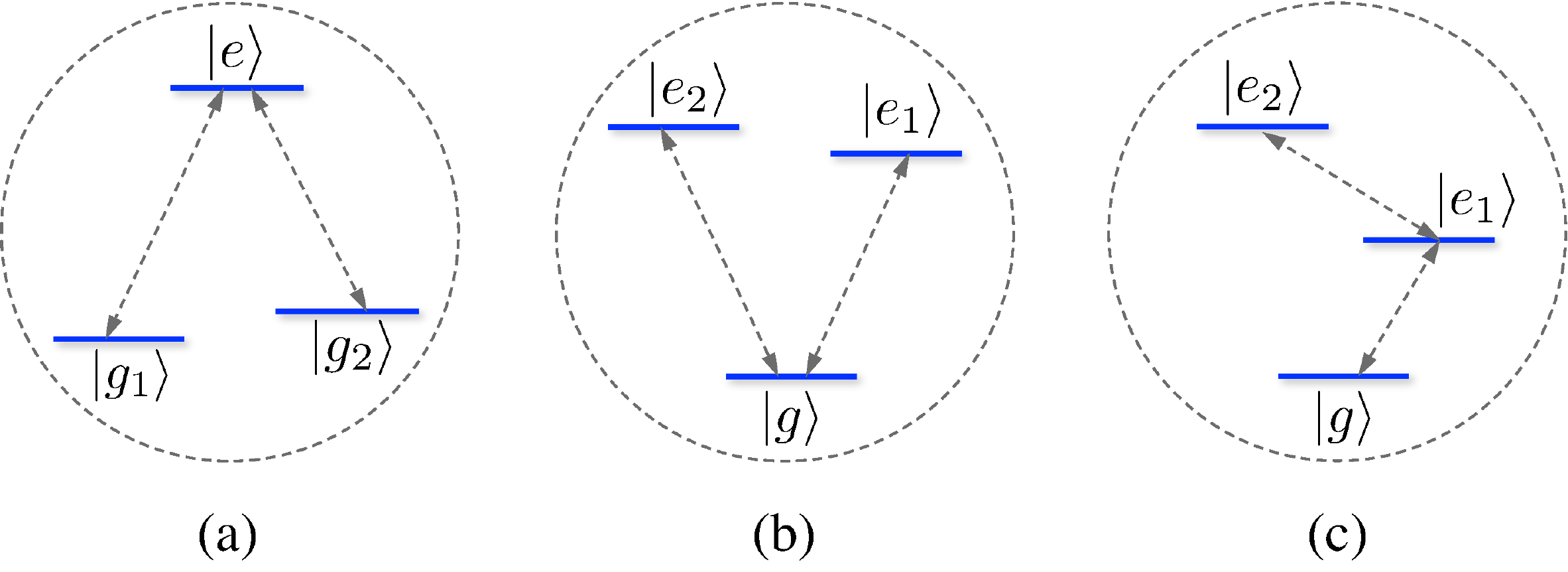}
\par\end{centering}

\protect\caption{(Color online) Three types of three-level systems, (a) $\Lambda$-type
(b) V-type (c) $\Xi$-type.}

\label{fig-3LS}
\end{figure}

We consider a $\Lambda$-type system {[}see Fig.\,\ref{fig-3LS}(a){]},
which is described by the Hamiltonian
\begin{equation}
\hat{H}_{S}=\sum_{n=1}^{3}E_{n}|E_{n}\rangle\langle E_{n}|,
\end{equation}
 where we denote the three energy levels by $|E_{n}\rangle=|g_{1}\rangle,\,|g_{2}\rangle$
(the lower two states) and $|E_{n}\rangle=|e\rangle$ (the highest
excited state) correspondingly, with eigen energy $E_{g_{1}},\, E_{g_{2}}$
and $E_{e}$.

Due to the interaction with the environment, there are two transitions
in the system, i.e., $|g_{1}\rangle\leftrightarrow|e\rangle$ and
$|g_{2}\rangle\leftrightarrow|e\rangle$. Here we use the lowering
and raising operators to represent these two transitions, denoted
as $\hat{\varphi}_{i}^{-}:=|g_{i}\rangle\langle e|$ and $\hat{\varphi}_{i}^{+}:=|e\rangle\langle g_{i}|$
respectively for $i=1,\,2$. And we denote the energy difference of
each transition as $\varepsilon_{i}:=E_{e}-E_{g_{i}}$.

The two heat baths are modeled as collections of boson modes, described
by the Hamiltonian
\begin{equation}
\hat{H}_{B}=\sum_{k_{L}}\omega_{k_{L}}\hat{b}_{k_{L}}^{\dagger}\hat{b}_{k_{L}}+\sum_{k_{R}}\omega_{k_{R}}\hat{b}_{k_{R}}^{\dagger}\hat{b}_{k_{R}}.
\end{equation}
 The two transitions $\hat{\varphi}_{i}^{\pm}$ of the $\Lambda$-type
system are coupled to both the two heat baths, which are in their
own equilibrium thermal states $\rho_{B,\alpha}^{\mathrm{th}}={\cal Z}_{\alpha}^{-1}\exp[-\sum\omega_{k_{\alpha}}\hat{b}_{k_{\alpha}}^{\dagger}\hat{b}_{k_{\alpha}}/T_{\alpha}]$
with temperatures $T_{\alpha}$ for $\alpha=L,\, R$ respectively.
The interaction Hamiltonian reads 
\begin{align}
\hat{H}_{SB}= & \hat{\varphi}_{1}^{+}\cdot(\hat{B}_{1,L}+\hat{B}_{1,R})+\hat{\varphi}_{1}^{-}\cdot(\hat{B}_{1,L}^{\dagger}+\hat{B}_{1,R}^{\dagger})\nonumber \\
+ & \hat{\varphi}_{2}^{+}\cdot(\hat{B}_{2,L}+\hat{B}_{2,R})+\hat{\varphi}_{2}^{-}\cdot(\hat{B}_{2,L}^{\dagger}+\hat{B}_{2,R}^{\dagger}),\label{eq:H_SB}
\end{align}
where 
\[
\hat{B}_{i,\alpha}=\sum_{k_{\alpha}}g_{i,k_{\alpha}}\hat{b}_{k_{\alpha}},\quad\alpha=L,R\text{ and }i=1,2
\]
is the collective operator of bath-$\alpha$ coupled to transition-$i$. 

To derive a master equation for this open quantum system, we apply
the Born-Markovian approximation in the interaction picture (without
the secular approximation) \cite{breuer_theory_2002},
\begin{equation}
\dot{\rho}=-\int_{0}^{\infty}ds\,\mathrm{Tr}_{B}\big[\hat{H}_{SB}(t),\,[\hat{H}_{SB}(t-s),\,\rho(t)\otimes\rho_{B}]\big].\label{eq:BMA}
\end{equation}
And we obtain the following time-dependent master equation in the
interaction picture (the detailed calculation is presented in \ref{sec:DerivationME}),
\begin{align}
\dot{\rho}= & \sum_{i,j=1}^{2}e^{i\Delta_{ij}t}\Big\{\Gamma_{ji}^{+}(\varepsilon_{j})\big[\hat{\varphi}_{i}^{+},\,\rho\hat{\varphi}_{j}^{-}\big]+\Gamma_{ji}^{+}(\varepsilon_{i})\big[\hat{\varphi}_{i}^{+}\rho,\,\hat{\varphi}_{j}^{-}\big]\Big\}\nonumber \\
 & +e^{-i\Delta_{ij}t}\Big\{\Gamma_{ij}^{-}(\varepsilon_{j})\big[\hat{\varphi}_{i}^{-},\,\rho\hat{\varphi}_{j}^{+}\big]+\Gamma_{ij}^{-}(\varepsilon_{i})\big[\hat{\varphi}_{i}^{-}\rho,\,\hat{\varphi}_{j}^{+}\big]\Big\},\label{eq:ME}
\end{align}
where $\Delta_{ij}:=\varepsilon_{i}-\varepsilon_{j}$, and
\begin{align}
\Gamma_{ij}^{-}(\omega) & :=\frac{1}{2}\gamma_{ij}^{(L)}(\omega)[N_{L}(\omega)+1]+\frac{1}{2}\gamma_{ij}^{(R)}(\omega)[N_{R}(\omega)+1],\nonumber \\
\Gamma_{ij}^{+}(\omega) & :=\frac{1}{2}\gamma_{ij}^{(L)}(\omega)N_{L}(\omega)+\frac{1}{2}\gamma_{ij}^{(R)}(\omega)N_{R}(\omega),\label{eq:dis-rates}
\end{align}
are called the dissipation rates. Here $N_{\alpha}(\omega):=[\exp(\omega/T_{\alpha})-1]^{-1}$
is the Planck distribution. And $\gamma_{ij}^{(\alpha)}(\omega)$
is the coupling spectrum of bath-$\alpha$, which is defined as 
\begin{equation}
\gamma_{ij}^{(\alpha)}(\omega):=2\pi\sum_{k_{\alpha}}g_{i,k_{\alpha}}^{*}g_{j,k_{\alpha}}\delta(\omega-\omega_{k_{\alpha}})=[\gamma_{ji}^{(\alpha)}(\omega)]^{*}.
\end{equation}

We should notice that besides the \emph{individual spectrum} $\gamma_{ii}(\omega)$
for each transition, we also have the \emph{cross spectrums} \cite{wei_time_1994},
$\gamma_{12}(\omega)=[\gamma_{21}(\omega)]^{*}$, which describe the
interference effect between the two transitions \cite{cardimona_steady-state_1982,scully_quantum_1997,kiffner_chapter_2010}. 

There is a relation between the cross spectrums $\gamma_{ij}^{(\alpha)}(\omega)$
and the individual spectrums $\gamma_{ii}^{(\alpha)}(\omega)$, shown
as follows (see the proof in \ref{sec:Correlated-spectrum}),
\begin{equation}
|\gamma_{12}^{(\alpha)}(\omega)|^{2}=f_{\alpha}(\omega)\cdot\gamma_{11}^{(\alpha)}(\omega)\gamma_{22}^{(\alpha)}(\omega),\label{eq:correlate-spectrum}
\end{equation}
where $f_{\alpha}(\omega)$ is a weight factor and $0\le f_{\alpha}(\omega)\le1$.
If $f_{\alpha}(\omega)=0$, it means there is no interference between
transitions. In some special cases, we have $f_{\alpha}(\omega)=1$,
which means the interference effect achieves the maximum. In general
cases, the concrete form of $f_{\alpha}(\omega)$ depends on the form
of coupling strength $g_{i,k_{\alpha}}$ in specific physical systems
\cite{mccutcheon_long-lived_2009,nalbach_quantum_2010} (see also
the example in \ref{sec:Correlated-spectrum}). For example, in quantum
optics, the weight factor $f_{\alpha}(\omega)$ is equivalent to the
$p$ parameter which measures the angle between the transition dipole
moments \cite{zhu_quantum-mechanical_1995,kozlov_inducing_2006,kiffner_chapter_2010,tscherbul_long-lived_2014,tscherbul_partial_2015}.

\subsection{Secular approximation}

We can understand the physical meaning of each summation term in the
master equation (\ref{eq:ME}) as a second order process. In fact,
due to the interaction with the environment, the system first absorbs
an energy quanta from the environment through the transition $\hat{\varphi}_{i}^{+}$,
and then immediately followed by emitting an energy quanta through
transition $\hat{\varphi}_{j}^{-}$ back to the environment. And the
reversed process also happens, namely, the system first emits an energy
quanta through transition $\hat{\varphi}_{i}^{-}$, and then absorbs
through transition $\hat{\varphi}_{j}^{+}$.

If these two successive energy exchange processes experience the same
transition ($i=j$), the total process is stable. While if these two
processes experience two different transitions ($i\neq j$), which
usually have different energies, the total process, with an oscillating
factor $\exp(i\Delta_{ij}t)$, is often considered to be not stable,
which will averagely vanish to zero after oscillations for several
periods $\Delta t\sim h/\Delta_{ij}$ \cite{jing_breakdown_2009}.
We call them cross transitions, and these terms results from the interference
of the two transitions. It was believed that such cross transition
terms would have no effect after a long enough oscillating time $t\gg\hbar/\left|\Delta_{ij}\right|$
\cite{breuer_theory_2002,jing_breakdown_2009}, and they do not contribute
to the steady state ($t\rightarrow\infty$).

For this reason, the secular approximation is often applied to Eq.\,(\ref{eq:ME}),
and only the terms with $\Delta_{ij}=\varepsilon_{i}-\varepsilon_{j}=0$
are remained. Then we obtain a time-independent Lindblad master equation
\cite{breuer_theory_2002,wu_quantum_2011,vogl_criticality_2012,yin_nonequilibrium_2014},
\begin{equation}
\dot{\rho}=\sum_{i=1}^{2}\Gamma_{ii}^{+}(\varepsilon_{i})\big(2\hat{\varphi}_{i}^{+}\rho\hat{\varphi}_{i}^{-}-\hat{\varphi}_{i}^{-}\hat{\varphi}_{i}^{+}\rho-\rho\hat{\varphi}_{i}^{-}\hat{\varphi}_{i}^{+}\big)+\Gamma_{ii}^{-}(\varepsilon_{i})\big(2\hat{\varphi}_{i}^{-}\rho\hat{\varphi}_{i}^{+}-\hat{\varphi}_{i}^{+}\hat{\varphi}_{i}^{-}\rho-\rho\hat{\varphi}_{i}^{+}\hat{\varphi}_{i}^{-}\big).\label{eq:ME-secular}
\end{equation}
 We should notice that all the cross transition terms are omitted
by the secular approximation in this master equation.

Here we make some clarification on our terminology in this paper.
When we say ``secular approximation'', we mean the omission of the
cross transition terms we mentioned above with energy difference $\Delta_{ij}=\varepsilon_{i}-\varepsilon_{j}\neq0$.
When we say ``rotating-wave approximation (RWA)'', we mean the omission
of double creation or annihilation terms, for example, in the derivation
of Jaynes-Cummings coupling from dipole interaction \cite{orszag_quantum_2000}.
The interaction Hamiltonian we used in Eq.\,(\ref{eq:H_SB}) is usually
obtained via RWA in real physics system.

For the master equation (\ref{eq:ME-secular}) with secular approximation,
the equations for the diagonal and off-diagonal terms of $\rho$ are
decoupled \cite{breuer_theory_2002,wu_quantum_2011,vogl_criticality_2012,yin_nonequilibrium_2014},
and we can obtain a rate equation only involving the populations $\overline{n}_{i}:=\langle g_{i}|\rho|g_{i}\rangle$
and $\overline{n}_{e}:=\langle e|\rho|e\rangle$ of each energy level
as follows,
\begin{align}
\dot{\overline{n}}_{1}= & 2\Gamma_{11}^{-}(\varepsilon_{1})\overline{n}_{e}-2\Gamma_{11}^{+}(\varepsilon_{1})\overline{n}_{1},\label{eq:rate}\\
\dot{\overline{n}}_{2}= & 2\Gamma_{22}^{-}(\varepsilon_{2})\overline{n}_{e}-2\Gamma_{22}^{+}(\varepsilon_{2})\overline{n}_{2}.\nonumber 
\end{align}
We notice that $\overline{n}_{1}+\overline{n}_{2}+\overline{n}_{e}=1$.

Setting $\dot{\overline{n}}_{i}=0$ in the rate equation (\ref{eq:rate}),
we obtain the steady population of the open quantum system, that is,
\begin{equation}
\frac{\overline{n}_{1}}{\overline{n}_{e}}=\frac{\Gamma_{11}^{-}(\varepsilon_{1})}{\Gamma_{11}^{+}(\varepsilon_{1})},\qquad\frac{\overline{n}_{2}}{\overline{n}_{e}}=\frac{\Gamma_{22}^{-}(\varepsilon_{2})}{\Gamma_{22}^{+}(\varepsilon_{2})}.\label{eq:secular-ss}
\end{equation}
We see that no matter whether the environment is in equilibrium, we
always obtain a steady state of a diagonal form $\rho_{s}=\sum P_{n}|n\rangle\langle n|$,
and all the off-diagonal terms of $\rho_{s}$ vanish. Thus, using
the master equation (\ref{eq:ME-secular}) with secular approximation,
we always obtain a solution where no quantum coherence is left after
a long-time evolution.

Here we make some clarification about the concept of ``quantum coherence''
we mention in this paper. Intuitively, people usually say there is
quantum coherence when there are some nonzero off-diagonal terms in
the density matrix, but for any density matrix, we can always make
it diagonalized in a certain basis. Thus it seems that the concept
of quantum coherence is not free of representation \cite{baumgratz_quantifying_2014}.
However, the energy representation is distinctive from others. The
canonical thermal state is diagonal only in the energy representation.
Therefore, in this paper, when we say there is quantum coherence,
we mean that the density matrix $\rho$ of the system has some non-zero
off-diagonal terms in the energy representation \cite{scully_quantum_1997}.

\section{Steady quantum coherence}

In this section, we take into account the interference between the
transitions, and thus we do not apply the secular approximation. We
show that non-vanishing quantum coherence can exist in the non-equilibrium
steady state of the $\Lambda$-type system when $T_{L}\neq T_{R}$.
Then we study the condition for the existence of steady quantum coherence.

\subsection{Steady state equation and numerical result}

Notice that the master equation (\ref{eq:ME}) without secular approximation
is time-independent in Schr\"odinger picture \cite{li_long-term_2014},
that is,
\begin{align}
\dot{\rho}=i[\rho,\,\hat{H}_{S}+\hat{H}_{\mathrm{c}}]+ & \sum_{i,j=1}^{2}[\Gamma_{ji}^{+}(\varepsilon_{i})+\Gamma_{ji}^{+}(\varepsilon_{j})]\big(\hat{\varphi}_{i}^{+}\rho\hat{\varphi}_{j}^{-}-\frac{1}{2}\{\rho,\,\hat{\varphi}_{j}^{-}\hat{\varphi}_{i}^{+}\}_{+}\big)\label{eq:ME-S}\\
 & +[\Gamma_{ij}^{-}(\varepsilon_{i})+\Gamma_{ij}^{-}(\varepsilon_{j})]\big(\hat{\varphi}_{i}^{-}\rho\hat{\varphi}_{j}^{+}-\frac{1}{2}\{\rho,\,\hat{\varphi}_{j}^{+}\hat{\varphi}_{i}^{-}\}_{+}\big),\nonumber 
\end{align}
where 
\[
\hat{H}_{\mathrm{c}}=\frac{1}{2i}\sum_{i,j=1}^{2}[\Gamma_{ji}^{+}(\varepsilon_{i})-\Gamma_{ji}^{+}(\varepsilon_{j})]\cdot\hat{\varphi}_{j}^{-}\hat{\varphi}_{i}^{+}+[\Gamma_{ij}^{-}(\varepsilon_{i})-\Gamma_{ij}^{-}(\varepsilon_{j})]\cdot\hat{\varphi}_{j}^{+}\hat{\varphi}_{i}^{-}
\]
 can be regarded as the non-diagonal Lamb shift resulted from interference
between transitions \cite{kiffner_chapter_2010}. Notice that Eq.\,(\ref{eq:ME-S})
has a modified Lindblad form \cite{lindblad_generators_1976}. From
this master equation, we obtain the dynamics of the population expectations
on the three levels of the $\Lambda$-type system, $\overline{n}_{i}:=\langle g_{i}|\rho|g_{i}\rangle$
and $\overline{n}_{e}:=\langle e|\rho|e\rangle$, as follows,
\begin{align}
\dot{\overline{n}}_{1}= & 2\Gamma_{11}^{-}(\varepsilon_{1})\overline{n}_{e}-2\Gamma_{11}^{+}(\varepsilon_{1})\overline{n}_{1}-\Gamma_{12}^{+}(\varepsilon_{2})\overline{\tau}_{12}-\Gamma_{21}^{+}(\varepsilon_{2})\overline{\tau}_{21},\label{eq:steady}\\
\dot{\overline{n}}_{2}= & 2\Gamma_{22}^{-}(\varepsilon_{2})\overline{n}_{e}-2\Gamma_{22}^{+}(\varepsilon_{2})\overline{n}_{2}-\Gamma_{12}^{+}(\varepsilon_{1})\overline{\tau}_{12}-\Gamma_{21}^{+}(\varepsilon_{1})\overline{\tau}_{21},\nonumber \\
\dot{\overline{\tau}}_{12}= & [\Gamma_{21}^{-}(\varepsilon_{1})\overline{n}_{e}-\Gamma_{21}^{+}(\varepsilon_{1})\overline{n}_{1}]+[\Gamma_{21}^{-}(\varepsilon_{2})\overline{n}_{e}-\Gamma_{21}^{+}(\varepsilon_{2})\overline{n}_{2}]+i\Delta_{12}\overline{\tau}_{12}-[\Gamma_{11}^{+}(\varepsilon_{1})+\Gamma_{22}^{+}(\varepsilon_{2})]\overline{\tau}_{12}.\nonumber 
\end{align}
Here we denote $\hat{\tau}_{12}:=|g_{1}\rangle\langle g_{2}|$, and
we have $\overline{\tau}_{12}=\langle g_{2}|\rho|g_{1}\rangle=\rho_{21}$. 

\begin{figure}
\begin{centering}
\includegraphics[width=0.9\textwidth]{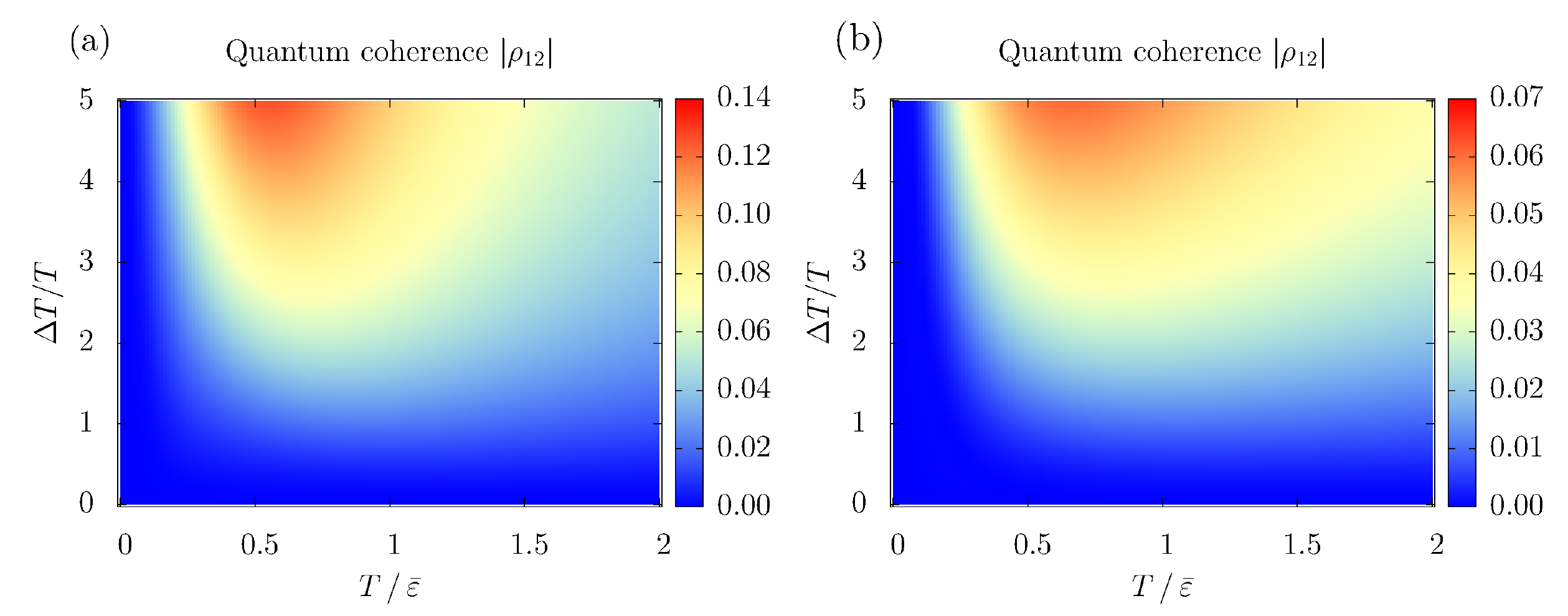}
\par\end{centering}

\protect\caption{(Color online) Numerical result for $|\rho_{12}|$ in the non-equilibrium
steady state of (a) $\Lambda$-type (b) V-type system. Here we set
$T_{L}=T$ and $T_{R}=T+\Delta T$. And we set $\overline{\varepsilon}=\frac{1}{2}(\varepsilon_{1}+\varepsilon_{2})=1$
as the energy unit. We set the coupling strengths to be $\gamma_{11}^{(L)}(\omega)=0.01,\,\gamma_{11}^{(R)}(\omega)=0.02$
and $\gamma_{22}^{(L)}(\omega)=0.02,\,\gamma_{22}^{(R)}(\omega)=0.01$.
The cross spectrums are set to be real, i.e., $\gamma_{12}^{(\alpha)}(\omega)=\gamma_{21}^{(\alpha)}(\omega)=[f_{\alpha}(\omega)]^{\frac{1}{2}}\cdot[\gamma_{11}^{(\alpha)}(\omega)\gamma_{22}^{(\alpha)}(\omega)]^{\frac{1}{2}}$
for $\alpha=L,R$, and we set the weight factors as $f_{L}(\omega)=f_{R}(\omega)=1$.
For the $\Lambda$-type system (a), we set the energy difference as
$\Delta_{12}=\varepsilon_{1}-\varepsilon_{2}=0.01$, while for the
V-type system (b) we set $\Delta_{12}=\varepsilon_{1}-\varepsilon_{2}=-0.01$.}

\label{fig-coh}
\end{figure}

We see that the dynamics of the populations $\overline{n}_{i}$ and
$\overline{n}_{g}$ are not decoupled from that of the off-diagonal
terms $\rho_{12/21}$, which is different from the case with secular
approximation Eq.\,(\ref{eq:rate}) where we obtained a closed system
of equations only about the populations. Thus $\rho_{12/21}$ may
not be zero even in the steady state after long time evolution. And
unlike previous study, this steady quantum coherence is not resulted
from the state degeneracy \cite{kozlov_inducing_2006}.

Setting $\dot{\overline{n}}_{i}=\dot{\overline{\tau}}_{12}=0$ in
Eq.\,(\ref{eq:steady}), we obtain the steady state of the open quantum
system by solving the linear equations, i.e., 
\begin{align}
-2\Gamma_{11}^{-}(\varepsilon_{1})\overline{n}_{e}= & -2\Gamma_{11}^{+}(\varepsilon_{1})\overline{n}_{1}-\Gamma_{12}^{+}(\varepsilon_{2})\overline{\tau}_{12}-\Gamma_{21}^{+}(\varepsilon_{2})\overline{\tau}_{21},\nonumber \\
-2\Gamma_{22}^{-}(\varepsilon_{2})\overline{n}_{e}= & -2\Gamma_{22}^{+}(\varepsilon_{2})\overline{n}_{2}-\Gamma_{12}^{+}(\varepsilon_{1})\overline{\tau}_{12}-\Gamma_{21}^{+}(\varepsilon_{1})\overline{\tau}_{21},\nonumber \\
-[\Gamma_{21}^{-}(\varepsilon_{1})+\Gamma_{21}^{-}(\varepsilon_{2})]\overline{n}_{e}= & -\Gamma_{21}^{+}(\varepsilon_{1})\overline{n}_{1}-\Gamma_{21}^{+}(\varepsilon_{2})\overline{n}_{2}-[\Gamma_{11}^{+}(\varepsilon_{1})+\Gamma_{22}^{+}(\varepsilon_{2})-i\Delta_{12}]\overline{\tau}_{12},\nonumber \\
-[\Gamma_{12}^{-}(\varepsilon_{1})+\Gamma_{12}^{-}(\varepsilon_{2})]\overline{n}_{e}= & -\Gamma_{12}^{+}(\varepsilon_{1})\overline{n}_{1}-\Gamma_{12}^{+}(\varepsilon_{2})\overline{n}_{2}-[\Gamma_{11}^{+}(\varepsilon_{1})+\Gamma_{22}^{+}(\varepsilon_{2})+i\Delta_{12}]\overline{\tau}_{21}.
\end{align}
This is a set of algebraic linear equations for $\overline{n}_{1},\,\overline{n}_{2},\,\overline{\tau}_{12},\,\overline{\tau}_{21}$,
and the determinant of the coefficient matrix is
\begin{align}
\det= & [\Gamma_{11}^{+}(\varepsilon_{1})+\Gamma_{22}^{+}(\varepsilon_{2})]^{2}\Big[4\Gamma_{11}^{+}(\varepsilon_{1})\Gamma_{22}^{+}(\varepsilon_{2})-4\Re\mathrm{e}[\Gamma_{12}^{+}(\varepsilon_{2})\Gamma_{21}^{+}(\varepsilon_{1})]+\Delta_{12}^{2}\Big]\nonumber \\
 & -\Big[2\Im\mathrm{m}[\Gamma_{12}^{+}(\varepsilon_{2})\Gamma_{21}^{+}(\varepsilon_{1})]+\Delta_{12}[\Gamma_{11}^{+}(\varepsilon_{1})-\Gamma_{22}^{+}(\varepsilon_{2})]\Big]^{2}.\label{eq:Det}
\end{align}
For the case when the weight factors $f_{L/R}(\omega)$ are real,
we have that $\Gamma_{12}^{+}(\omega)=\Gamma_{21}^{+}(\omega)$ are
also real, and
\begin{equation}
\det=4[\Gamma_{11}^{+}(\varepsilon_{1})+\Gamma_{22}^{+}(\varepsilon_{2})]^{2}[\Gamma_{11}^{+}(\varepsilon_{1})\Gamma_{22}^{+}(\varepsilon_{2})-\Gamma_{12}^{+}(\varepsilon_{2})\Gamma_{21}^{+}(\varepsilon_{1})]+4\Delta_{12}^{2}\Gamma_{11}^{+}(\varepsilon_{1})\Gamma_{22}^{+}(\varepsilon_{2}).\label{eq:det}
\end{equation}
If $\det=0$, the system does not have a unique steady state, and
the long time behavior depends on the initial state. For example,
if we choose $\Delta_{12}=0$, $\gamma_{ii}^{(L)}(\varepsilon_{1})=\gamma_{ii}^{(L)}(\varepsilon_{2})=\gamma_{ii}^{(R)}(\varepsilon_{1})=\gamma_{ii}^{(R)}(\varepsilon_{2})$
and $f_{L/R}(\varepsilon_{i})=1$ for $i=1,2$, we can check from
Eq.\,(\ref{eq:det}) that we would always get $\det=0$, no matter
how much the temperatures $T_{L/R}$ take. Indeed this is just the
case in the study of spontaneous decay-induced coherences \cite{zhu_quantum-mechanical_1995,kozlov_inducing_2006,kiffner_chapter_2010}.
But this condition is usually too strong for practical physical systems.
Especially, here we care more about non-degenerated cases $\Delta_{12}\neq0$.
Indeed, in usual cases $\det=0$ seldom happen, or it only has several
discretely distributed solutions, which have measure 0 in the parameter
space of $\varepsilon_{1},\varepsilon_{2},T_{L},T_{R},\gamma_{ij}^{(\alpha)}(\omega)$,
unless the spectrums $\gamma_{ij}^{(\alpha)}(\omega)$ have some very
novel structures. Therefore, for most practical cases, the steady
state is unique and does not depend on the initial state.

We present a numerical result for the steady state in Fig.\,\ref{fig-coh}(a),
where we set $T_{L}=T$ and $T_{R}=T+\Delta T$. The steady solution
is unique and does not depend on the initial condition (We checked
this numerically for parameters in Fig.\,\ref{fig-coh}). We see
that $|\rho_{12}|=|\overline{\tau}_{21}|$ does not vanish in the
steady state, that is, there exist steady quantum coherence in the
non-equilibrium system. And the amplitude of $|\rho_{12}|$ is not
a negligibly small value. Here we have set $f_{L}(\omega)=f_{R}(\omega)=1$
to achieve the maximum effect of interference between transitions.
If we have $f_{\alpha}(\omega)<1$, the steady quantum coherence is
suppressed, but still keeps nonzero. This is different from some previous
studies about the noise-induced coherence in quantum optical setups,
where the weight factor $f_{\alpha}(\omega)$ (or the $p$ parameter
measuring the angle between the transition dipole moments in quantum
optics) must take its maximum $1$ \cite{zhu_quantum-mechanical_1995,kozlov_inducing_2006}.

Moreover, we should notice that when the lower temperature $T_{L}=T$
is given, the nonzero off-diagonal term $|\rho_{12}|$ increases with
the temperature difference $\Delta T$. Besides, when the lower temperatures
$T_{L}$ is quite high or approaches the zero temperature, $|\rho_{12}|$
becomes small but keeps nonzero. We also need to emphasize that the
two states, $|g_{1}\rangle$ and $|g_{2}\rangle$, which have steady
quantum coherence between them, do not need to be degenerated, which
is different from what has been studied in previous literatures \cite{rahav_heat_2012}.

We need to emphasize that the steady quantum coherence we obtained
above is not resulted from the ``decoherence-free subspace'' (DFS)
\cite{duan_preserving_1997,zanardi_noiseless_1997,lidar_decoherence-free_1998},
and there is no DFS in this model. For a system with DFS, the states
in the DFS are decoupled from the environment, and they evolve unitarily.
The quantum coherence protected in the DFS is determined by the initial
state, and there does not exist a unique steady state in such systems.
However, in our model, the steady state, which is obtained from Eq.\,(\ref{eq:steady}),
is unique and does not depend on the initial state, thus there is
no DFS here and the steady coherence in this model is not resulted
from the DFS.

Now we show that when the temperature difference $\Delta T$ decreases
to zero exactly, i.e., when we return to case of an equilibrium heat
bath with a single temperature, the steady quantum coherence also
vanishes completely. In this case, we have $T_{L}=T_{R}:=T$, and
$N_{L}(\omega)=N_{R}(\omega):=N(\omega)$. From Eq.\,(\ref{eq:dis-rates})
we see that the dissipation rates $\Gamma_{ij}^{\pm}(\omega)$ satisfy
the following relation of \emph{micro-reversibility} \cite{bergmann_new_1955,dallot_average-ion_1998},
\begin{equation}
\frac{\Gamma_{ij}^{+}(\omega)}{\Gamma_{ij}^{-}(\omega)}=e^{-\omega/T},\label{eq:micro-Rever}
\end{equation}
which leads to the detailed balance and the equilibrium distribution.
With the help of the above relation, we can verify that 
\begin{equation}
\frac{\overline{n}_{i}}{\overline{n}_{g}}=\frac{\Gamma_{ii}^{+}(\varepsilon_{i})}{\Gamma_{ii}^{-}(\varepsilon_{i})}=e^{-\varepsilon_{i}/T},\quad\rho_{12}=\rho_{21}^{*}=0,\label{eq:res-eq}
\end{equation}
is exactly the steady solution of Eq.\,(\ref{eq:steady}). Thus,
in the equilibrium case, all the off-diagonal terms vanish to zero
after long-time evolution, and this steady solution is same with the
result Eq.\,(\ref{eq:secular-ss}) obtained from the rate equation
with secular approximation. That means, the secular approximation
is consistent in the case of equilibrium environment.

\subsection{Condition for steady quantum coherence}

Now we give the condition for the existence of steady quantum coherence.
As we will see below, non-equilibrium is a necessary but not sufficient
condition for the existence of steady quantum coherence.

To guarantee there is no steady quantum coherence in the steady state
$\rho_{12}=\rho_{21}^{*}=0$, the necessary condition is 
\begin{equation}
\Gamma_{21}^{+}(\varepsilon_{1})\big[\frac{\Gamma_{21}^{-}(\varepsilon_{1})}{\Gamma_{21}^{+}(\varepsilon_{1})}-\frac{\Gamma_{11}^{-}(\varepsilon_{1})}{\Gamma_{11}^{+}(\varepsilon_{1})}\big]+\Gamma_{21}^{+}(\varepsilon_{2})\big[\frac{\Gamma_{21}^{-}(\varepsilon_{2})}{\Gamma_{21}^{+}(\varepsilon_{2})}-\frac{\Gamma_{22}^{-}(\varepsilon_{2})}{\Gamma_{22}^{+}(\varepsilon_{2})}\big]=0.\label{eq:zero-condition}
\end{equation}
This can be derived directly by putting $\overline{\tau}_{12}=0$
into Eq.\,(\ref{eq:steady}). Thus it is the necessary condition
to guarantee zero quantum coherence in the steady state. Moreover,
if the steady solution is unique {[}see Eq.\,(\ref{eq:Det}) and
the discussion below{]}, the above condition is also sufficient {[}we
just need to verify that the trial solution $\overline{\tau}_{12}=0$
and Eq.\,(\ref{eq:secular-ss}) are consistent with the above condition
(\ref{eq:zero-condition}), and the uniqueness of the solution guarantees
the sufficiency{]}. Notice that, as a special case, in the equilibrium
case we mentioned above, the dissipation rates automatically satisfy
that the two terms in condition (\ref{eq:zero-condition}) both equal
to zero, which roots from the micro-reversibility Eq.\,(\ref{eq:micro-Rever}).

Besides the equilibrium case, there is another case that the condition
(\ref{eq:zero-condition}) still hold even for non-equilibrium environment.
That is, when the coupling spectrums $\gamma_{ij}^{(\alpha)}$ satisfy
the following relation 
\begin{equation}
\frac{\gamma_{21}^{(L)}(\varepsilon_{1})}{\gamma_{21}^{(R)}(\varepsilon_{1})}=\frac{\gamma_{11}^{(L)}(\varepsilon_{1})}{\gamma_{11}^{(R)}(\varepsilon_{1})},\qquad\frac{\gamma_{21}^{(L)}(\varepsilon_{2})}{\gamma_{21}^{(R)}(\varepsilon_{2})}=\frac{\gamma_{22}^{(L)}(\varepsilon_{2})}{\gamma_{22}^{(R)}(\varepsilon_{2})}.\label{eq:NEQ-condition}
\end{equation}
It can be verified from Eq.\,(\ref{eq:dis-rates}) that the above
relation also guarantees 
\begin{equation}
\frac{\Gamma_{21}^{-}(\varepsilon_{1})}{\Gamma_{21}^{+}(\varepsilon_{1})}-\frac{\Gamma_{11}^{-}(\varepsilon_{1})}{\Gamma_{11}^{+}(\varepsilon_{1})}=0,\quad\frac{\Gamma_{21}^{-}(\varepsilon_{2})}{\Gamma_{21}^{+}(\varepsilon_{2})}-\frac{\Gamma_{22}^{-}(\varepsilon_{2})}{\Gamma_{22}^{+}(\varepsilon_{2})}=0.
\end{equation}
Thus the condition (\ref{eq:zero-condition}) is satisfied to give
rise to vanishing quantum coherence. If the relation (\ref{eq:NEQ-condition})
is not satisfied, usually we obtain nonzero $\rho_{12/21}$ in the
steady state.

Now we discuss the physical meaning of the relation (\ref{eq:NEQ-condition}).
We will show that the relation (\ref{eq:NEQ-condition}) implies that
the two transitions couple to the two heat baths with the same strength
proportion. 

\begin{figure}
\begin{centering}
\includegraphics[width=0.66\textwidth]{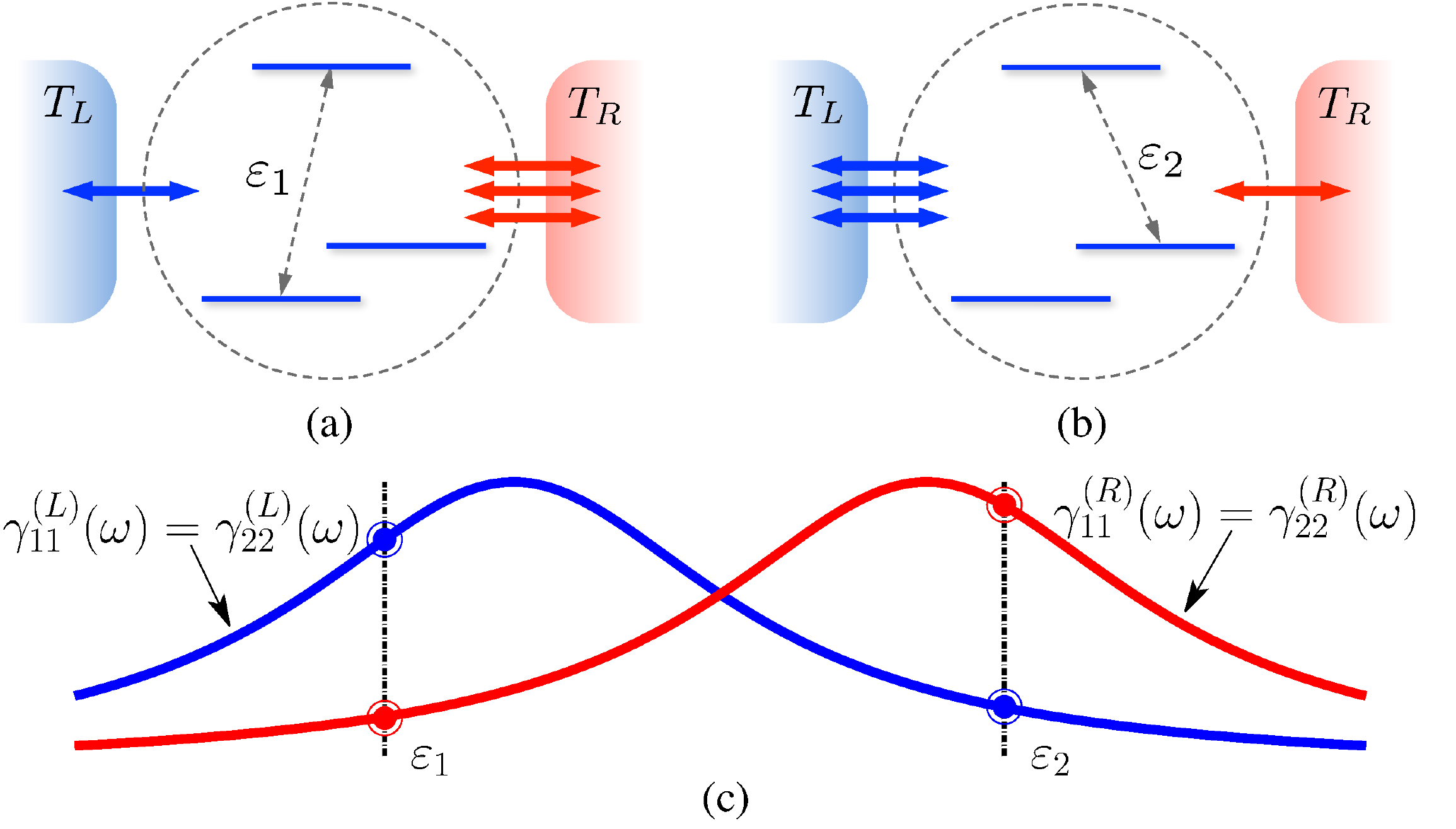}
\par\end{centering}

\protect\caption{(Color online) (a) transition-1 couples more strongly to bath-$R$,
$\gamma_{11}^{(L)}(\varepsilon_{1})<\gamma_{11}^{(R)}(\varepsilon_{1})$.
(b) transition-2 couples more strongly to bath-$L$, $\gamma_{22}^{(L)}(\varepsilon_{2})>\gamma_{22}^{(R)}(\varepsilon_{2})$.
(c) An example of spectrums $\gamma_{ii}^{(\alpha)}(\omega)$ to support
such configurations of (a) and (b).}

\label{fig-couple}
\end{figure}

Remember that we have a relation about the cross spectrum, $|\gamma_{12}^{(\alpha)}(\omega)|^{2}=f_{\alpha}(\omega)\cdot\gamma_{11}^{(\alpha)}(\omega)\gamma_{22}^{(\alpha)}(\omega)$
{[}Eq.\,(\ref{eq:correlate-spectrum}){]}. Here we consider a simple
case that $f_{L}(\omega)=f_{R}(\omega)>0$. In this case, the module
square of Eq.\,(\ref{eq:NEQ-condition}) gives the following relation,
\begin{equation}
\frac{\gamma_{11}^{(L)}(\varepsilon_{1})}{\gamma_{11}^{(R)}(\varepsilon_{1})}=\frac{\gamma_{22}^{(L)}(\varepsilon_{1})}{\gamma_{22}^{(R)}(\varepsilon_{1})},\qquad\frac{\gamma_{11}^{(L)}(\varepsilon_{2})}{\gamma_{11}^{(R)}(\varepsilon_{2})}=\frac{\gamma_{22}^{(L)}(\varepsilon_{2})}{\gamma_{22}^{(R)}(\varepsilon_{2})}.\label{eq:prop}
\end{equation}
 That means, for the two transitions, their coupling strengths with
the two heat baths must be of the same proportion, so as to guarantee
vanishing quantum coherence.

Here we give an example for the spectrums that the above relation
(\ref{eq:prop}) is broken. As demonstrated in Fig.\,\ref{fig-couple}(a,
b), transition-1 couples to bath-$R$ more strongly than bath-$L$
{[}$\gamma_{11}^{(L)}(\varepsilon_{1})<\gamma_{11}^{(R)}(\varepsilon_{1})${]},
but on contrary transition-2 couples to bath-$L$ more strongly than
bath-$R$ {[}$\gamma_{22}^{(L)}(\varepsilon_{2})>\gamma_{22}^{(R)}(\varepsilon_{2})${]}.
We can check that the above strength proportion relation Eq.\,(\ref{eq:prop})
is violated, and so is the condition (\ref{eq:zero-condition}). In
this case, there exists non-vanishing steady quantum coherence between
the two lower states $|g_{1/2}\rangle$. In the above numerical result
{[}Fig.\,\ref{fig-coh}(a){]}, the coupling strengths $\gamma_{ij}^{(\alpha)}(\varepsilon_{i})$
are chosen just in this way. Such configuration can be realized by
imposing coupling spectrums of proper shapes as shown in Fig.\,\ref{fig-couple}(c).

\section{V-type and $\Xi$-type systems}

We have shown that for a $\Lambda$-type system in a non-equilibrium
environment, the interference between transitions can give rise to
non-vanishing steady quantum coherence. In this section, we consider
the case of V-type and $\Xi$-type systems.

We should notice that the master equation (\ref{eq:ME-S}) we derived
for a $\Lambda$-type system is also valid for V-type and $\Xi$-type
systems, as long as we properly redefine the raising and lowering
operators $\hat{\varphi}_{i}^{\pm}$ for the two transitions. 

For a V-type system, we define $\hat{\varphi}_{i}^{-}:=|g\rangle\langle e_{i}|$
and $\hat{\varphi}_{i}^{+}:=|e_{i}\rangle\langle g|$ {[}see the notations
in Fig.\,\ref{fig-3LS}(b){]}. With these operators for the transitions,
we can obtain a master equation having the same form with Eq.\,(\ref{eq:ME-S}).
The steady state equation for the V-type system is
\begin{align}
\dot{\overline{n}}_{1}= & 2\Gamma_{11}^{+}(\varepsilon_{1})\overline{n}_{g}-2\Gamma_{11}^{-}(\varepsilon_{1})\overline{n}_{1}-\Gamma_{12}^{-}(\varepsilon_{2})\overline{\tau}_{12}-\Gamma_{21}^{-}(\varepsilon_{2})\overline{\tau}_{21},\\
\dot{\overline{n}}_{2}= & 2\Gamma_{22}^{+}(\varepsilon_{2})\overline{n}_{g}-2\Gamma_{22}^{-}(\varepsilon_{2})\overline{n}_{2}-\Gamma_{12}^{-}(\varepsilon_{1})\overline{\tau}_{12}-\Gamma_{21}^{-}(\varepsilon_{1})\overline{\tau}_{21},\nonumber \\
\dot{\overline{\tau}}_{12}= & [\Gamma_{21}^{+}(\varepsilon_{1})\overline{n}_{g}-\Gamma_{21}^{-}(\varepsilon_{1})\overline{n}_{1}]+[\Gamma_{21}^{+}(\varepsilon_{2})\overline{n}_{g}-\Gamma_{21}^{-}(\varepsilon_{2})\overline{n}_{2}]+i\Delta_{12}\overline{\tau}_{12}-[\Gamma_{11}^{-}(\varepsilon_{1})+\Gamma_{22}^{-}(\varepsilon_{2})]\overline{\tau}_{12}.\nonumber 
\end{align}
Here we denote $\overline{n}_{i}:=\langle e_{i}|\rho|e_{i}\rangle$,
$\overline{n}_{g}:=\langle g|\rho|g\rangle$, $\hat{\tau}_{12}:=|e_{1}\rangle\langle e_{2}|$
and $\overline{\tau}_{12}=\langle e_{2}|\rho|e_{1}\rangle$. We see
again that the populations $\overline{n}_{i}$ and $\overline{n}_{e}$
also depend on the value of off-diagonal terms $\rho_{12}$ and $\rho_{21}$.
As we discussed above, in the steady state, we can also obtain nonzero
quantum coherence in the steady state. We show the numerical result
of the steady quantum coherence in the V-type system in Fig.\,\ref{fig-coh}(b).
Notice that the maximum value of $|\rho_{12}|$ in the V-type system
is much smaller than that in the $\Lambda$-type system {[}see Fig.\,\ref{fig-coh}(a){]},
because the steady quantum coherence in the V-type system exists between
excited energy levels, which possess much less populations.

For the $\Xi$-type system, we define $\hat{\varphi}_{1}^{+}:=|e_{1}\rangle\langle g|$,
$\hat{\varphi}_{2}^{+}:=|e_{2}\rangle\langle e_{1}|$, and $\hat{\varphi}_{i}^{-}=[\hat{\varphi}_{i}^{+}]^{\dagger}$
for the two transitions {[}see the notations in Fig.\,\ref{fig-3LS}(c){]}.
The master equation of the $\Xi$-type system still has the form of
Eq.\,(\ref{eq:ME-S}), while the steady state equation is
\begin{align}
\dot{\overline{n}}_{1}= & 2[\Gamma_{11}^{+}(\varepsilon_{1})\overline{n}_{g}-\Gamma_{11}^{-}(\varepsilon_{1})\overline{n}_{1}],\\
\dot{\overline{n}}_{2}= & 2[\Gamma_{22}^{+}(\varepsilon_{2})\overline{n}_{1}-\Gamma_{22}^{-}(\varepsilon_{2})\overline{n}_{2}].\nonumber 
\end{align}
Here we denote $\overline{n}_{i}:=\langle e_{i}|\rho|e_{i}\rangle$
and $\overline{n}_{g}:=\langle g|\rho|g\rangle$. Different from the
case of V-type and $\Lambda$-type systems, the populations alone
form a closed set of equations and do not depend on any off-diagonal
terms. More precisely speaking, the interference between transitions
do not affect the time evolution of the populations. Thus, for the
$\Xi$-type system, there is always no quantum coherence left in the
steady state, and this result is consistent with secular approximation
{[}Eq.\,(\ref{eq:secular-ss}){]}.

\section{Physical meaning of quantum coherence}

In this section, we demonstrate an example to study the physical meaning
of the quantum coherence. In this example, we will see that the quantum
coherence between the energy eigenstates exactly reflects the non-equilibrium
flux inside the composite system \cite{li_long-term_2014}.

\begin{figure}
\begin{centering}
\includegraphics[width=0.85\textwidth]{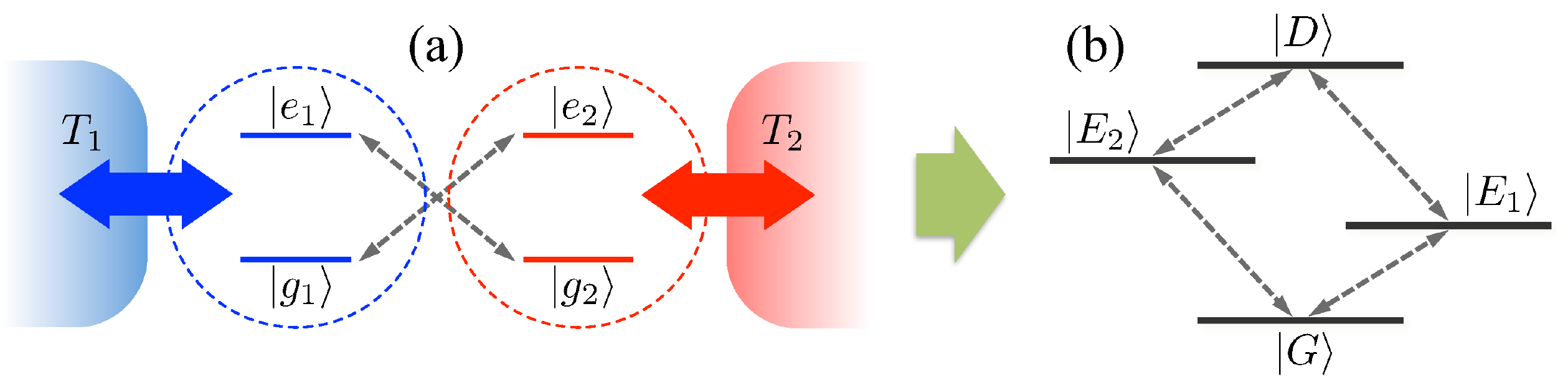}
\par\end{centering}

\protect\caption{(Color online) (a) Two coupled TLSs. The two TLSs exchange energy
with each other. We denote the bare states of each TLS by $|e_{i}\rangle,\,|g_{i}\rangle$,
and the eigenstates of the composite system are $|G\rangle,\,|E_{i}\rangle,\,|D\rangle$.
The interaction with the environment gives rise to transitions between
the four eigenstates, as shown by arrows in (b), where $|E_{2}\rangle\leftrightarrow|G\rangle\leftrightarrow|E_{1}\rangle$
can be regarded as a V-type structure, and $|E_{2}\rangle\leftrightarrow|D\rangle\leftrightarrow|E_{1}\rangle$
can be regarded as a $\Lambda$-type structure.}

\label{fig-2TLS}
\end{figure}

We consider a composite system of two coupled two-level-systems (TLSs),
which is described by 
\begin{equation}
\hat{H}_{S}=\frac{1}{2}\omega_{1}\hat{\sigma}_{1}^{z}+\frac{1}{2}\omega_{2}\hat{\sigma}_{2}^{z}+g(\hat{\sigma}_{1}^{+}\hat{\sigma}_{2}^{-}+\hat{\sigma}_{1}^{-}\hat{\sigma}_{2}^{+}),
\end{equation}
where $\hat{\sigma}_{i}^{+}:=|e_{i}\rangle\langle g_{i}|,\,\hat{\sigma}_{i}^{-}:=|g_{i}\rangle\langle e_{i}|,\,\hat{\sigma}_{i}^{z}:=|e_{i}\rangle\langle e_{i}|-|g_{i}\rangle\langle g_{i}|$,
and $|e_{1,2}\rangle,\,|g_{1,2}\rangle$ are bare states of each TLS
{[}see Fig.\,\ref{fig-2TLS}(a){]}. And this Hamiltonian can be diagonalized
as $\hat{H}_{S}=\sum_{n}E_{n}|E_{n}\rangle\langle E_{n}|$. The eigen
energies and the corresponding eigen states are \cite{liao_quantum_2011}
\begin{alignat}{2}
E_{G} & =-\frac{1}{2}\overline{\Omega}, & \quad|G\rangle & =|g_{1}g_{2}\rangle,\label{eq:eigen}\\
E_{1} & =-\frac{1}{2}\sqrt{\Delta^{2}+4g^{2}}, & \quad|E_{1}\rangle & =\sin\frac{\theta}{2}|e_{1}g_{2}\rangle-\cos\frac{\theta}{2}|g_{1}e_{2}\rangle,\nonumber \\
E_{2} & =\frac{1}{2}\sqrt{\Delta^{2}+4g^{2}}, & \quad|E_{2}\rangle & =\cos\frac{\theta}{2}|e_{1}g_{2}\rangle+\sin\frac{\theta}{2}|g_{1}e_{2}\rangle,\nonumber \\
E_{D} & =\frac{1}{2}\overline{\Omega}, & \quad|D\rangle & =|e_{1}e_{2}\rangle,\nonumber 
\end{alignat}
where $\overline{\Omega}:=(\omega_{1}+\omega_{2})/2,\,\Delta:=\omega_{1}-\omega_{2}$
and $\cot\theta=\Delta/2g$.

Each TLS couples with an independent heat bath via interaction $\hat{H}_{SB}=\hat{H}_{SB}^{(1)}+\hat{H}_{SB}^{(2)}$,
where
\begin{equation}
\hat{H}_{SB}^{(\alpha)}=\sum_{k_{\alpha}}g_{k_{\alpha}}\,\hat{\sigma}_{\alpha}^{+}\hat{b}_{k_{\alpha}}+g_{k_{\alpha}}^{*}\,\hat{\sigma}_{\alpha}^{-}\hat{b}_{k_{\alpha}}^{\dagger}.
\end{equation}
Such interaction gives rise to a transition structure as shown in
Fig.\,\ref{fig-2TLS}(b) \cite{liao_quantum_2011,li_synchro-thermalization_2014}.
We can regard the transition $|E_{2}\rangle\leftrightarrow|G\rangle\leftrightarrow|E_{1}\rangle$
as a V-type structure, and $|E_{2}\rangle\leftrightarrow|D\rangle\leftrightarrow|E_{1}\rangle$
as a $\Lambda$-type structure, and there exists quantum interference
between the transitions. Such a non-equilibrium system can be realized
in present experiments by, for example, interacting superconducting
qubits \cite{chen_quantum_2012}, or double quantum dots \cite{li_cooling_2011,cai_entropy_2014},
etc.

Now we consider the non-equilibrium flux flowing across, for example,
the TLS-$1$. The dynamics of the population of TLS-$1$ $\overline{n}_{1}:=\langle e_{1}|\rho|e_{1}\rangle$
can be obtained by the Heisenberg equation,
\begin{align}
\dot{\overline{n}}_{1} & =\frac{d}{dt}\langle\hat{\sigma}_{1}^{z}\rangle=-i\langle[\hat{\sigma}_{1}^{z},\,\hat{H}_{S}+\hat{H}_{SB}+\hat{H}_{B}]\rangle\\
 & =-i\langle[\hat{\sigma}_{1}^{z},\,\hat{H}_{S}]\rangle-i\langle[\hat{\sigma}_{1}^{z},\,\hat{H}_{SB}]\rangle:=J_{1-2}+J_{1-B_{1}},\nonumber 
\end{align}
where $J_{1-2}:=-i\langle[\hat{\sigma}_{1}^{z},\,\hat{H}_{S}]\rangle$
means the internal flux between the two TLSs, while $J_{1-B_{1}}$
means the flux flowing between TLS-$1$ and bath-$1$. We have
\begin{align*}
J_{1-2} & =-i2g\langle\hat{\sigma}_{1}^{+}\hat{\sigma}_{2}^{-}-\hat{\sigma}_{1}^{-}\hat{\sigma}_{2}^{+}\rangle=-i2g\mathrm{Tr}\Big[\rho(|e_{1}g_{2}\rangle\langle g_{1}e_{2}|-|g_{1}e_{2}\rangle\langle e_{1}g_{2}|)\Big]\\
 & =-i2g\mathrm{Tr}\Big[\rho(|E_{2}\rangle\langle E_{1}|-|E_{1}\rangle\langle E_{2}|)\Big]=4g\Im\mathrm{m}\langle E_{1}|\rho|E_{2}\rangle.
\end{align*}

The above calculation is completed with the help of Eq.\,(\ref{eq:eigen}).
The above equation clearly shows that the non-equilibrium flux between
the two TLSs $J_{1-2}$ is exactly reflected by the imaginary part
of the quantum coherence term $\langle E_{1}|\rho|E_{2}\rangle$ \cite{li_long-term_2014}.
If the two heat baths have the same temperature, there is no net heat
transfer, and the steady quantum coherence automatically vanishes,
which is just consistent with what we have discussed above.

Therefore, the quantum coherence exactly reflects the non-equilibrium
flux inside a composite system. If the steady quantum coherence is
missed, the internal flux inside the composite system would also be
omitted improperly, which leads to an unphysical conclusion that there
is no net flux between the local sites even when the two heat baths
have different temperatures.

\section{Conclusion}

In this paper, we studied the steady state of a three-level system
in a non-equilibrium environment, which consists of two heat baths
with different temperatures. We find that for the $\Lambda$-type
and V-type systems, the interference between transitions can give
rise to non-vanishing steady quantum coherence, if the two transitions
couple to the two heat baths with different proportions of coupling
strengths. And the amount of the steady quantum coherence increases
with the temperature difference of the two heat baths. If the two
heat baths have the same temperature, all the quantum coherence vanishes
and returns to the equilibrium case. These transition structures are
quite common in natural and artificial quantum systems. The non-equilibrium
environment can be implemented via current noises with different effective
temperatures in quantum circuits \cite{chen_quantum_2012,venturelli_minimal_2013},
or electron leads with different chemical potentials \cite{li_cooling_2011}.

The interference between transitions play an essential role in the
steady quantum coherence. But it was often omitted by secular approximation
in previous literatures. We showed that indeed the secular approximation
is consistent in the case of equilibrium environment, but for non-equilibrium
environments, that would lead to the neglect of the steady quantum
coherence.

We also show that the quantum coherence has a clear physical meaning,
i.e., it exactly reflects the internal non-equilibrium flux inside
a composite system, which is an important characterization of non-equilibrium
systems.

Notice that many current investigations about non-equilibrium quantum
thermodynamics are based on the rate equation of the energy level
populations like Eq.\,(\ref{eq:rate}), which does not include the
quantum coherence. Our result implies that some further refinement,
which takes into account the quantum coherence, should be made to
the present study of non-equilibrium quantum thermodynamics \cite{cai_entropy_2014}.

\section*{Acknowledgement}

This work was supported by the National Natural Science Foundation
of China (Grant No.\,11121403), the National 973-program (Grant No.\,2012CB922104
and No.\,2014CB921403), and Postdoctoral Science Foundation of China
No.\,2013M530516. S.-W. Li wants to thank D. Z. Xu and Z. H. Wang
for helpful discussion.

\appendix

\section{Derivation of the master equation\label{sec:DerivationME}}

We show some detailed calculation of the derivation for the master
equation. Denoting $\hat{\mathsf{B}}_{i}=\hat{B}_{i,L}+\hat{B}_{i,R}$,
we rewrite the interaction Hamiltonian Eq.\,(\ref{eq:H_SB}) as $\hat{H}_{SB}:=\sum_{i}\hat{\varphi}_{i}^{+}\cdot\hat{\mathsf{B}}_{i}+\hat{\varphi}_{i}^{-}\cdot\hat{\mathsf{B}}_{i}^{\dagger}$.
In the interaction picture of $\hat{H}_{S}+\hat{H}_{B}$, we have
\begin{equation}
\hat{H}_{SB}(t)=\sum_{i}\hat{\varphi}_{i}^{+}(t)\cdot\hat{\mathsf{B}}_{i}(t)+\hat{\varphi}_{i}^{-}(t)\cdot\hat{\mathsf{B}}_{i}^{\dagger}(t):=V^{+}(t)+V^{-}(t).
\end{equation}
 Put it into Eq.\,(\ref{eq:BMA}) of $\dot{\rho}(t)$, and we have
\begin{equation}
\dot{\rho}=-\int_{0}^{\infty}ds\,\mathrm{Tr}_{B}\Big\{\big[\hat{V}^{+}(t),\,[\hat{V}^{-}(t-s),\,\rho(t)\otimes\rho_{B}]\big]+\big[\hat{V}^{-}(t),\,[\hat{V}^{+}(t-s),\,\rho(t)\otimes\rho_{B}]\big]\Big\}.
\end{equation}
The above equation is expanded as follows
\begin{align*}
\dot{\rho}= & -\int_{0}^{\infty}ds\sum_{i,j=1}^{2}\Big\{ e^{i(\varepsilon_{i}-\varepsilon_{j})t}\cdot\hat{\varphi}_{i}^{+}\hat{\varphi}_{j}^{-}\rho\cdot e^{i\varepsilon_{j}s}\langle\hat{\mathsf{B}}_{i}(t)\hat{\mathsf{B}}_{j}^{\dagger}(t-s)\rangle-e^{i(\varepsilon_{i}-\varepsilon_{j})t}\cdot\hat{\varphi}_{i}^{+}\rho\hat{\varphi}_{j}^{-}\cdot e^{i\varepsilon_{j}s}\langle\hat{\mathsf{B}}_{j}^{\dagger}(t-s)\hat{\mathsf{B}}_{i}(t)\rangle\\
 & -e^{i(\varepsilon_{i}-\varepsilon_{j})t}\cdot\hat{\varphi}_{j}^{-}\rho\hat{\varphi}_{i}^{+}\cdot e^{i\varepsilon_{j}s}\langle\hat{\mathsf{B}}_{i}(t)\hat{\mathsf{B}}_{j}^{\dagger}(t-s)\rangle+e^{i(\varepsilon_{i}-\varepsilon_{j})t}\cdot\rho\hat{\varphi}_{j}^{-}\hat{\varphi}_{i}^{+}\cdot e^{i\varepsilon_{j}s}\langle\hat{\mathsf{B}}_{j}^{\dagger}(t-s)\hat{\mathsf{B}}_{i}(t)\rangle\\
\\
 & +e^{-i(\varepsilon_{i}-\varepsilon_{j})t}\cdot\hat{\varphi}_{i}^{-}\hat{\varphi}_{j}^{+}\rho\cdot e^{-i\varepsilon_{j}s}\langle\hat{\mathsf{B}}_{i}^{\dagger}(t)\hat{\mathsf{B}}_{j}(t-s)\rangle-e^{-i(\varepsilon_{i}-\varepsilon_{j})t}\cdot\hat{\varphi}_{i}^{-}\rho\hat{\varphi}_{j}^{+}\cdot e^{-i\varepsilon_{j}s}\langle\hat{\mathsf{B}}_{j}(t-s)\hat{\mathsf{B}}_{i}^{\dagger}(t)\rangle\\
 & -e^{-i(\varepsilon_{i}-\varepsilon_{j})t}\cdot\hat{\varphi}_{j}^{+}\rho\hat{\varphi}_{i}^{-}\cdot e^{-i\varepsilon_{j}s}\langle\hat{\mathsf{B}}_{i}^{\dagger}(t)\hat{\mathsf{B}}_{j}(t-s)\rangle+e^{-i(\varepsilon_{i}-\varepsilon_{j})t}\cdot\rho\hat{\varphi}_{j}^{+}\hat{\varphi}_{i}^{-}\cdot e^{-i\varepsilon_{j}s}\langle\hat{\mathsf{B}}_{j}(t-s)\hat{\mathsf{B}}_{i}^{\dagger}(t)\rangle\Big\}.
\end{align*}
We apply the Born approximation that the two heat bath always stay
at their canonical thermal state with temperature $T_{\alpha}$ respectively.
Here we show the calculation of two terms as demonstration,
\begin{align}
 & \int_{0}^{\infty}ds\, e^{i\varepsilon s}\langle\hat{\mathsf{B}}_{i}(t)\hat{\mathsf{B}}_{j}^{\dagger}(t-s)\rangle=\int_{0}^{\infty}ds\, e^{i\varepsilon s}\Big(\langle\hat{B}_{i,L}(t)\hat{B}_{j,L}^{\dagger}(t-s)\rangle+\langle\hat{B}_{i,R}(t)\hat{B}_{j,R}^{\dagger}(t-s)\rangle\Big)\nonumber \\
= & \int_{0}^{\infty}ds\int\frac{d\omega}{2\pi}\, e^{i(\varepsilon-\omega)s}\Big(\gamma_{ji}^{(L)}(\omega)[N_{L}(\omega)+1]+\gamma_{ji}^{(R)}(\omega)[N_{R}(\omega)+1]\Big)\nonumber \\
= & \frac{1}{2}\gamma_{ji}^{(L)}(\varepsilon)[N_{L}(\varepsilon)+1]+\frac{1}{2}\gamma_{ji}^{(R)}(\varepsilon)[N_{R}(\varepsilon)+1]:=\Gamma_{ji}^{-}(\varepsilon),
\end{align}
\begin{align}
 & \int_{0}^{\infty}ds\, e^{i\varepsilon s}\langle\hat{\mathsf{B}}_{i}^{\dagger}(t)\hat{\mathsf{B}}_{j}(t-s)\rangle=\int_{0}^{\infty}ds\, e^{i\varepsilon s}\Big(\langle\hat{B}_{i,L}^{\dagger}(t)\hat{B}_{j,L}(t-s)\rangle+\langle\hat{B}_{i,R}^{\dagger}(t)\hat{B}_{j,R}(t-s)\rangle\Big)\nonumber \\
= & \int_{0}^{\infty}ds\int\frac{d\omega}{2\pi}\, e^{i(\varepsilon-\omega)s}\Big(\gamma_{ij}^{(L)}(\omega)N_{L}(\omega)+\gamma_{ij}^{(R)}(\omega)N_{R}(\omega)\Big)\nonumber \\
= & \frac{1}{2}\gamma_{ij}^{(L)}(\varepsilon)N_{L}(\varepsilon)+\frac{1}{2}\gamma_{ij}^{(R)}(\varepsilon)N_{R}(\varepsilon):=\Gamma_{ij}^{+}(\varepsilon),
\end{align}
where we define the coupling spectrum with bath-$\alpha$ as $\gamma_{ij}^{(\alpha)}(\omega):=2\pi\sum g_{i,k_{\alpha}}^{*}g_{j,k_{\alpha}}\delta(\omega-\omega_{k_{\alpha}})=[\gamma_{ji}^{(\alpha)}(\omega)]^{*}$.
Here we utilized the formula
\begin{equation}
\int_{0}^{\infty}ds\, e^{i(\varepsilon-\omega)s}=\pi\delta(\varepsilon-\omega)+i\mathbf{P}\frac{1}{\varepsilon-\omega},
\end{equation}
and omitted the principle integral terms. Then we obtain the master
equation as
\begin{align*}
\dot{\rho}=-\sum_{i,j=1}^{2}\Big\{ & \Gamma_{ji}^{-}(\varepsilon_{j})e^{i\Delta_{ij}t}\hat{\varphi}_{i}^{+}\hat{\varphi}_{j}^{-}\rho-\Gamma_{ji}^{+}(\varepsilon_{j})e^{i\Delta_{ij}t}\hat{\varphi}_{i}^{+}\rho\hat{\varphi}_{j}^{-}-\Gamma_{ji}^{-}(\varepsilon_{j})e^{i\Delta_{ij}t}\hat{\varphi}_{j}^{-}\rho\hat{\varphi}_{i}^{+}+\Gamma_{ji}^{+}(\varepsilon_{j})e^{i\Delta_{ij}t}\rho\hat{\varphi}_{j}^{-}\hat{\varphi}_{i}^{+}\\
+\Gamma_{ij}^{+}(\varepsilon_{j}) & e^{-i\Delta_{ij}t}\hat{\varphi}_{i}^{-}\hat{\varphi}_{j}^{+}\rho-\Gamma_{ij}^{-}(\varepsilon_{j})e^{-i\Delta_{ij}t}\hat{\varphi}_{i}^{-}\rho\hat{\varphi}_{j}^{+}-\Gamma_{ij}^{+}(\varepsilon_{j})e^{-i\Delta_{ij}t}\hat{\varphi}_{j}^{+}\rho\hat{\varphi}_{i}^{-}+\Gamma_{ij}^{-}(\varepsilon_{j})e^{-i\Delta_{ij}t}\rho\hat{\varphi}_{j}^{+}\hat{\varphi}_{i}^{-}\Big\},
\end{align*}
where $\Delta_{ij}:=\varepsilon_{i}-\varepsilon_{j}$. In Schr\"odinger
picture, we can write down the master equation in the following time-independent
Lindblad-like form, 
\begin{align}
\dot{\rho}=i[\rho,\,\hat{H}_{S}]+\sum_{i,j=1}^{2}\Big\{ & \Gamma_{ji}^{+}(\varepsilon_{j})\cdot\big[\hat{\varphi}_{i}^{+},\,\rho\hat{\varphi}_{j}^{-}\big]+\Gamma_{ji}^{+}(\varepsilon_{i})\cdot\big[\hat{\varphi}_{i}^{+}\rho,\,\hat{\varphi}_{j}^{-}\big]\\
+ & \Gamma_{ij}^{-}(\varepsilon_{j})\cdot\big[\hat{\varphi}_{i}^{-},\,\rho\hat{\varphi}_{j}^{+}\big]+\Gamma_{ij}^{-}(\varepsilon_{i})\cdot\big[\hat{\varphi}_{i}^{-}\rho,\,\hat{\varphi}_{j}^{+}\big]\Big\},\nonumber 
\end{align}
or equivalently,
\begin{align}
\dot{\rho}=i[\rho,\,\hat{H}_{S}+\hat{H}_{\mathrm{c}}]+\sum_{i,j=1}^{2} & [\Gamma_{ji}^{+}(\varepsilon_{i})+\Gamma_{ji}^{+}(\varepsilon_{j})]\cdot\big(\hat{\varphi}_{i}^{+}\rho\hat{\varphi}_{j}^{-}-\frac{1}{2}\{\rho,\,\hat{\varphi}_{j}^{-}\hat{\varphi}_{i}^{+}\}_{+}\big)\\
+ & [\Gamma_{ij}^{-}(\varepsilon_{i})+\Gamma_{ij}^{-}(\varepsilon_{j})]\cdot\big(\hat{\varphi}_{i}^{-}\rho\hat{\varphi}_{j}^{+}-\frac{1}{2}\{\rho,\,\hat{\varphi}_{j}^{+}\hat{\varphi}_{i}^{-}\}_{+}\big),\nonumber 
\end{align}
where 
\begin{equation}
\hat{H}_{\mathrm{c}}=\sum_{i,j=1}^{2}\frac{\Gamma_{ji}^{+}(\varepsilon_{i})-\Gamma_{ji}^{+}(\varepsilon_{j})}{2i}\hat{\varphi}_{j}^{-}\hat{\varphi}_{i}^{+}+\frac{\Gamma_{ij}^{-}(\varepsilon_{i})-\Gamma_{ij}^{-}(\varepsilon_{j})}{2i}\hat{\varphi}_{j}^{+}\hat{\varphi}_{i}^{-}
\end{equation}
 can be regarded as the non-diagonal Lamb shift resulted from interference
between transitions \cite{kiffner_chapter_2010}.

\section{Cross spectrum\label{sec:Correlated-spectrum}}

Here we derive a relation of the cross spectrum. When we have two
transitions contacting with the same environment, we need three coupling
spectrums, i.e., two individual spectrum for each transition, $\gamma_{ii}(\omega)$,
and another cross spectrum \cite{wei_time_1994} for the cross transition,
$\gamma_{12}(\omega)$. These spectrums are defined as

\begin{align}
\gamma_{ii}(\omega) & :=2\pi\sum_{k}\left|g_{i,k}\right|^{2}\,\delta(\omega-\omega_{k}),\nonumber \\
\gamma_{12}(\omega) & :=2\pi\sum_{k}g_{1,k}^{*}\, g_{2,k}\,\delta(\omega-\omega_{k}).
\end{align}
From this definition, we have
\begin{align*}
\gamma_{11}(\omega)\gamma_{22}(\omega) & =4\pi^{2}\sum_{k,q}g_{1,k}^{*}g_{1,k}\delta(\omega-\omega_{k})\cdot g_{2,q}^{*}g_{2,q}\delta(\omega-\omega_{q}),\\
|\gamma_{12}(\omega)|^{2} & =4\pi^{2}\sum_{k,q}g_{1,k}^{*}g_{2,k}\delta(\omega-\omega_{k})\cdot g_{1,q}g_{2,q}^{*}\delta(\omega-\omega_{q}).
\end{align*}
From $|g_{1,k}g_{2,q}-g_{2,k}g_{1,q}|^{2}\ge0$, we have
\[
g_{1,k}^{*}g_{1,k}\cdot g_{2,q}^{*}g_{2,q}+g_{1,q}^{*}g_{1,q}\cdot g_{2,k}^{*}g_{2,k}\ge g_{1,k}^{*}g_{2,k}\cdot g_{1,q}g_{2,q}^{*}+g_{1,k}g_{2,k}^{*}\cdot g_{2,q}^{*}g_{2,q}.
\]
Thus, we have
\begin{align*}
\sum_{k,q}(g_{1,k}^{*}g_{1,k}\cdot g_{2,q}^{*}g_{2,q}+g_{1,q}^{*}g_{1,q} & \cdot g_{2,k}^{*}g_{2,k})\cdot\delta(\omega-\omega_{k})\delta(\omega-\omega_{q})\\
\ge\sum_{k,q}(g_{1,k}^{*}g_{2,k}\cdot g_{1,q}g_{2,q}^{*}+g_{1,k}g_{2,k}^{*} & \cdot g_{2,q}^{*}g_{2,q})\cdot\delta(\omega-\omega_{k})\delta(\omega-\omega_{q}).
\end{align*}
That is
\begin{equation}
\gamma_{11}(\omega)\gamma_{22}(\omega)\ge|\gamma_{12}(\omega)|^{2}.
\end{equation}

From the above proof we see that the equality holds if and only if
we have 
\begin{equation}
g_{1,k}\cdot g_{2,q}=g_{1,q}\cdot g_{2,k},\label{eq:gg=00003Dgg}
\end{equation}
 for any $k,\, q$ which satisfy $\omega_{k}=\omega_{q}$. 

If $k\rightarrow\omega_{k}$ is a one-to-one map, we have $\omega_{k}=\omega_{q}\,\Leftrightarrow\, k=q$.
Then the above relation holds and further we have 
\begin{equation}
\gamma_{11}(\omega)\gamma_{22}(\omega)=|\gamma_{12}(\omega)|^{2}.\label{eq:f=00003D1}
\end{equation}
 In more general cases, for example, the environment is the electromagnetic
field or the phonon field, with the index $\mathbf{k}$ as a vector,
the mode energy $\omega_{\mathbf{k}}$ has degeneracy in different
directions of $\mathbf{k}$. If the amplitudes $\left|\mathbf{k}\right|$
are the same, the modes with different directions have the same energy
$\omega_{\mathbf{k}}$. In these cases, if the coupling coefficients
$g_{i,\mathbf{k}}=g_{i,|\mathbf{k}|}$ only depend on the amplitude
$|\mathbf{k}|$, i.e., only depend on $\omega_{\mathbf{k}}$ equivalently,
the relation (\ref{eq:gg=00003Dgg}) still holds, and we still have
$\gamma_{11}(\omega)\gamma_{22}(\omega)=|\gamma_{12}(\omega)|^{2}$,
as used in many literatures \cite{liao_quantum_2011}. 

Except the above two cases, the equality (\ref{eq:f=00003D1}) usually
does not hold, and the general relation between the cross spectrum
$\gamma_{12}(\omega)$ and the individual spectrums $\gamma_{ii}(\omega)$
appear as \cite{mccutcheon_long-lived_2009,kiffner_chapter_2010,nalbach_quantum_2010}
\begin{equation}
|\gamma_{12}(\omega)|^{2}=f(\omega)\cdot\gamma_{11}(\omega)\gamma_{22}(\omega),
\end{equation}
 where $0\le f(\omega)\le1$ is a weight factor. In general cases,
the concrete form of $f(\omega)$ depends on the form of coupling
strength $g_{i,\mathbf{k}}$ in specific physical systems. 

For example, two two-level atoms stay at position $\mathbf{r}_{1}$
and $\mathbf{r}_{2}$ in the same electromagnetic field, and they
separate from each other for a distance $|\mathbf{r}_{1}-\mathbf{r}_{2}|:=d$.
Their coupling strengths with the electromagnetic field have a relation
$g_{1,\mathbf{k}}=g_{2,\mathbf{k}}\exp[i\mathbf{k}\cdot(\mathbf{r}_{2}-\mathbf{r}_{1})]$.
The light emitted from the two atoms can interfere with each other.
In this case, we can check that the weight factor has the form of
$f(\omega d/c)$. It varies with the distance $d$, and depends on
the dimensionality $D$ of the electromagnetic field, i.e.,
\begin{equation}
f(x)=\begin{cases}
\cos^{2}(x), & D=1,\\
{}[J_{0}(x)]^{2}, & D=2,\\
\text{sinc}^{2}(x), & D=3,
\end{cases}
\end{equation}
where $J_{0}(x)$ is the Bessel function of the first kind \cite{mccutcheon_long-lived_2009}.
When the two atoms are quite near to each other, their interference
effect achieve the maximum. When they are far from each other, their
interference effect quickly decays with the distance (for $D=2,\,3$).
Thus the weight function $f(\omega d/c)$ here describe the spatial
correlation of the environment. The situation is similar when we consider
the phonon bath \cite{nalbach_quantum_2010}.

\end{document}